# Quaternary compounds $Ag_2XYSe_4$ (X=Ba, Sr; Y=Sn, Ge) as novel potential thermoelectric materials

A. J. Hong*[a], C. L. Yuan*[a], and J. –M. Liu[b]

*Correspondence and requests for materials should be addressed to A.J.H. (email: 6312886haj@163.com) and C.L.Y. (email: clyuan@jxnu.edu.cn).

[a] *Jiangxi Key Laboratory of Nanomaterials and Sensors, School of Physics, Communication and Electronics, Jiangxi Normal University, Nanchang 330022, China*

[b] *Laboratory of Solid State Microstructures and Innovation Center of Advanced Microstructures, Nanjing University, Nanjing 210093, China*

**ABSTRACT**: Experimental results have shown that the quaternary compound $Cu_2ZnSnSe_4$ is an excellent thermoelectric material. This inspires us to seek the other quaternary compounds with similar chemical formula to $Cu_2ZnSnSe_4$ as thermoelectric materials. In this paper, we use the first-principles method to systematically explore the electronic and phonon structures, mechanical, thermal and thermoelectric properties of p- and n-type $Ag_2XYSe_4$ (X=Ba, Sr; Y=Sn, Ge). It is found that the ZT maximum for n-type $Ag_2SrGeSe_4$ can reach up to 1.22 at 900 K, and those for p-type $Ag_2SrSnSe_4$, $Ag_2SrGeSe_4$ and $Ag_2BaSnSe_4$ can reach up to 1.20, 1.13 and 1.12, respectively. Our work not only shows that $Ag_2XYSe_4$ (X=Ba, Sr; Y=Sn, Ge) are a kind of potential thermoelectric materials, but also can inspire more theoretical and experimental researches on thermoelectric properties of quaternary compounds.

## I. INTRODUCTION

During the last few decades, with the increasing problems of energy exhaustion and environment pollution, thermoelectric (TE) materials have already aroused people's widespread concern [1, 2], due to their special properties of directly converting heat to electricity and vice versa. One can use a dimensionless figure of merit, namely $ZT=S^2\sigma T/\kappa$, to assess the TE materials' conversion efficiency from heat to electricity [3, 4]. Herein, $S$ and $\sigma$ represent the Seebeck coefficient and electrical conductivity, $T$ and $\kappa$ are the temperature (in unit of Kelvin) and total thermal conductivity composed of electronic $\kappa_e$ and lattice $\kappa_L$ contributions. Thus, a good TE material requires high $S$ and $\sigma$ and low $\kappa$. However, this is hard to achieve due to the complex interdependences between $S$, $\sigma$ and $\kappa$ [5]. Recently, it was reported that achieving large band degeneracy [6] or localized resonant states [7] via energy band engineering do improve the TE transport properties in few existing TE materials [8]. Nevertheless, this strategy has great blindness because one is difficult to decide which dopant can induce beneficial change to electronic structure. Another common strategy is to reduce lattice thermal conductivity by low-dimension [9] or solid solution strategies [10, 11] that usually require very rigorous preparation conditions. Thus, seeking new TE materials with medium TE performance and then improving ZT value by tuning the carrier density remains to be a simple and feasible method [12, 13].

How to screen good thermoelectric materials through electronic structures has always been of concern to researchers. For example, Liu, etc. [14] propose the generalized material parameter $B^* \propto U^*E_g/\kappa_L$ as a screening criterion where the weighted mobility $U^*$, the total thermal conductivity $\kappa_L$, and the band gap $E_g$. Good TE material should have $B^*$ as high as possible. Thus, theoretically, wide bandgap materials should possess better thermoelectric performance than narrow bandgap materials. Unfortunately, the carrier density of wide bandgap materials is regulated nearly impossible to the optimal density. Due to adjustability in carrier density, materials with bandgap of less than 1.0 eV usually are as candidates for TE application. For example, star TE materials $Bi_2Te_3$, SnSe, PbS, PbSe, PbTe, $CoSb_3$, NbFeSb, $Mg_2Si$, $Mg_2Ge$ and $Mg_2Sn$ have band gaps of ~0.15 eV [15], ~0.86 eV [5], ~0.4 eV [16], ~0.29 eV [16],

~0.32 eV [16], ~0.22 eV [17], ~0.51eV [18], ~0.77 eV [19], ~0.74 eV [20] and ~0.35 eV [21], respectively.

Recently, quaternary compounds $Cu_2ZnSn(S/Se)_4$ (CZTSSe) are considered as promising photovoltaic absorber material due to high efficiency of about 12.6% [22]. It was reported that the band gaps of pure $Cu_2ZnSnSe_4$ (CZTSe) and $Cu_2ZnSnS_4$ (CZTS) were about 1.0 [23-25] and 1.5 eV [26, 27] close to ideal values of 1.3–1.4 eV for solar cell materials. By judging from band gaps, instead of CZTS, compound CZTSe seems to be a promising thermoelectric material due to smaller band gap suitable for carrier density regulation. Indeed, previous works have indicated that the ZT of ln-doped CZTSe reached up to 0.95 at 850 K [28] and that of modified CZTS was only 0.36 at 700K [29]. Therefore, we tried to replace Cu, Zn and Sn atoms of compound $Cu_2ZnSnSe_4$ with Ag, Ba/Sr and Sn/Ge atoms, constructing four new compounds $Ag_2XYSe_4$ (X=Ba, Sr; Y=Sn, Ge). The structures of $Ag_2BaGeSe_4$ and $Ag_2BaSnSe_4$ have been experimentally verified to be space group $I222$ (No. 23) [30] and the other two were theoretically predicted to be space group $I222$ [31], also. Obviously, $Ag_2XYSe_4$ structures possess lower symmetry than kesterite and stannite $Cu_2ZnSnSe_4$ belonging to space groups $I\bar{4}$ (No. 82) and $I\bar{4}2m$ (No. 121) [32], respectively. Low symmetry can favorite the strengthening anharmonic vibration of phonon, and thus the low symmetry materials such as SnSe, $Zn_4Sb_3$ and MgAgSb usually possess low lattice thermal conductivity [33]. Furthermore, it is expected $Ag_2XYSe_4$ have high electrical conductivity due to containing two Ag atoms in each primitive cell.

To the best of our knowledge, there have been few theoretical reports on the TE properties of $Ag_2XYSe_4$ (X=Ba, Sr; Y=Sn, Ge). Hence, in this work, we used density functional theory (DFT) with semi-classical Boltzmann equation to explore the TE property of p- and n-type $Ag_2XYSe_4$ (X=Ba, Sr; Y=Sn, Ge). It was showed that band gaps of $Ag_2BaGeSe_4$, $Ag_2BaSnSe_4$, $Ag_2SrGeSe_4$ and $Ag_2SrSnSe_4$ are 0.909 eV, 0.832 eV, 0.708 eV and 0.729 eV, respectively. In addition, their lattice thermal conductivities are 2.22 $Wm^{-1}K^{-1}$, 1.95 $Wm^{-1}K^{-1}$, 2.26 $Wm^{-1}K^{-1}$ and 2.00 $Wm^{-1}K^{-1}$ at 300 K that are obviously lower than 3.2 $Wm^{-1}K^{-1}$ of $Cu_2ZnSnSe_4$ [28]. Lattice thermal conductivity at high temperature for each compound is below 1.0 $Wm^{-1}K^{-1}$. It is predicted that the n-type Ag$_2$SrGeSe$_4$, having ZT maximum of 1.22 (at 900K), is the most excellent in the compounds Ag$_2$XYSe$_4$ (X=Ba, Sr; Y=Sn, Ge), and its p-type counterpart has a ZT maximum of 1.13. The ZT maximums of both p-type Ag$_2$BaSnSe$_4$ and Ag$_2$SrSnSe$_4$ are above unity. This paper reveals Ag$_2$XYSe$_4$ are a new family of promising candidates for TE materials and may pave a way for seeking TE materials with high the ZT values.

We organize the rest of this paper as follows: the detailed computational method and some important parameters for first-principles calculations are presented in Sec. II. The electronic structures, mechanical, thermal and thermoelectric properties are described and discussed in Sec. III. Finally, findings and conclusions are summarized in Sec. IV.

## II. THEORETICAL CALCULATION DETAILS

Firstly, structure optimization was performed using the generalized gradient approximation of Perdew, Burke, and Ernzerhof (GGA-PBE) [34] as the electronic exchange-correlation functional in VASP code. The plane-wave energy cutoff and Monckhorst-Pack $k$-point mesh were set to 500 eV and 10×10×10, and the total-energy and force convergences were $10^{-6}$ eV and $10^{-3}$ eV/Å.

Secondly, we used the full potential method with GGA-PBE modified Becke and Johnson potential scheme (mBJ) [35] to perform atomic coordinate optimization and then calculate electronic structures for all the compounds in Wien2k code [36]. In the whole calculation procedure, the total-energy and charge convergences were taken as 0.0001 eV and 0.0001 $e$. 3000 $k$-points in the whole first Brillouin Zone were sampled for calculations of such as band structure and density of states. Taking a very dense mesh of 20000 $k$-points, the electrical transport parameters such as the Seebeck coefficient, electrical conductivity and electronic thermal conductivity, were calculated using the semi-classical Boltzmann theory. In this theoretical framework, the electronic transport parameters can described as follows [37]:

$$\sigma_{\alpha\beta}(T,E_F) = \frac{1}{\Omega} \int \sigma_{\alpha\beta}(\varepsilon) \left[ -\frac{\partial f}{\partial \varepsilon} \right] d\varepsilon, \quad (1)$$

$$S_{\alpha\beta}(T,E_F) = (\sigma^{-1})_{l\alpha} \eta_{l\beta}, \quad (2)$$

$$\kappa_{\alpha\beta}(T, E_F) = \kappa_{\alpha\beta}^0 - T\eta_{\alpha m}(\sigma^{-1})_{nm}\eta_{m\beta}, \tag{3}$$

where $T$, $E_f$, $\Omega$, $f$ and $\varepsilon$ are the absolute temperature, Fermi energy, unit cell volume, Fermi-Dirac distribution function and energy for electronic state. Noted, $\kappa_{\alpha\beta}$ is the electronic thermal conductivity without the external electric field. The energy projected conductivity tensors, $\sigma_{\alpha\beta}$, can be expressed as the following form

$$\sigma_{\alpha\beta}(\varepsilon) = \frac{e^2\tau}{N}\sum_{i,k} v_\alpha v_\beta \frac{\delta(\varepsilon - \varepsilon_{i,k})}{d\varepsilon}. \tag{4}$$

Herein, $e$ is the charge of an electron, $\tau$ is the carrier relaxation time, $N$ is the total number of sampled $k$-points, $\varepsilon_{i,k}$ is the energy for the $k$th point at the $i$th band. The other physical parameters are given as follows:

$$\eta_{\alpha\beta}(T, E_f) = \frac{1}{eT\Omega}\int \sigma_{\alpha\beta}(\varepsilon)(\varepsilon - E_f)\left[-\frac{\partial f}{\partial \varepsilon}\right]d\varepsilon, \tag{5}$$

$$\kappa_{\alpha\beta}^0(T, E_f) = \frac{1}{e^2T\Omega}\int \sigma_{\alpha\beta}(\varepsilon)(\varepsilon - E_f)^2\left[-\frac{\partial f}{\partial \varepsilon}\right]d\varepsilon. \tag{6}$$

Subsequently, we used the slack's equation (see Eq. 7) to calculate the lattice thermal conductivity for all systems [38].

$$\kappa_L = A\frac{\Theta_D^3 V_{per}^{1/3}\overline{m}}{\gamma_a^2 n_{tot}^{2/3} T}. \tag{7}$$

Herein, A is a collection of physical constants with the value of about $3.04\times10^{-6}$, $\Theta_D$ is the Debye temperature, $V_{per}$ is the volume per atom, $\overline{m}$ is the average atom mass in the whole unit cell, $n_{tot}$ is the total number of all atoms in the primitive cell. $\gamma_a$ is the Grüneisen parameter for only acoustic phonons. Previous works [39-41] show lattice thermal conductivity attained by the Slack's equation is good agreement with experimental data. For the purpose of comparison, we also used compressive sensing lattice dynamics (CSLD) method [42] combined with Boltzmann transport equation (BTE) [43] to compute the lattice thermal conductivity. In addition, we used the Phonopy code [44] to calculated phonon spectrum.

### III. RESULTS AND DISCUSSION
**A Crystal structures and electronic structures**

Crystal structures of compounds $Ag_2XYSe_4$ have been verified experimentally or theoretically to belong to group I222 (No. 23). There are 16 and 8 atoms in the conventional and primitive cells, respectively (see Fig. 1). The lattice constants of conventional cells in this work and other literatures are summarized in Table I. It is seen that our results for Sr compounds are good agreement with other theoretical data. However, for Ba compounds, there is a little difference compared to the experimental data. Experimental and theoretical lattice constants are determined at room and zero temperatures respectively, thus the difference is perfectly acceptable. The cell volume of $Ag_2SrGeSe_4$ is smaller than that of $Ag_2SrSnSe_4$ due to the difference in the atomic radii of Ge and Sn, which is as the same as Ba compounds. Similarly, due to Ba radius smaller than Sr radius, the cell volumes of $Ag_2BaGeSe_4$ and $Ag_2BaSnSe_4$ are smaller than those of $Ag_2SrGeSe_4$ and $Ag_2SrSnSe_4$, respectively.

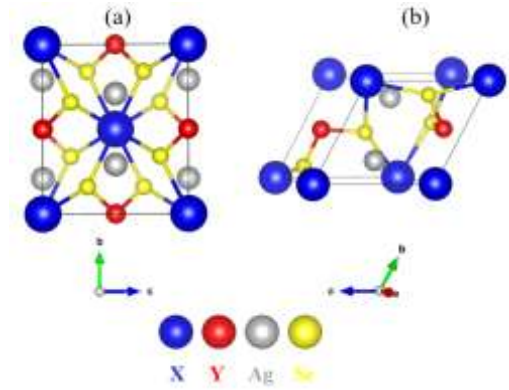

Fig. 1. (Color online) The (a) conventional (b) and primitive cells of $Ag_2XYSe_4$.

Calculated band structures and corresponding density of states (DOS) of the four compounds are showed in Fig. 2. Sn compounds have indirect band gaps of 0.765 eV ($Ag_2BaSnSe_4$) and 0.694 eV ($Ag_2SrSnSe_4$), of which valence band maximum (VBM) and conduction band minimum (CBM) are at the Z and Γ points respectively. However, the Ge compounds have direct band gaps of 0.870 eV ($Ag_2BaGeSe_4$) and 0.667 eV ($Ag_2SrGeSe_4$), with both CBM and VBM appearing at Z point. It is worth mentioning the difference between indirect and direct band gaps for each compound is very small. We have also noticed that other work [31] points $Ag_2BaGeSe_4$ and $Ag_2SrGeSe_4$ possess indirect band gaps. Perhaps, the full potential method with mBJ scheme was adopted in this work, which is different from the pseudopotential method, leading to this difference.

The PBE, mBJ and HSE06-hybrid (from Ref. [31]) band gaps for the four compounds are also listed in Table I. The mBJ band gaps, close to the HSE06-hybrid band gaps, are obviously larger than the PBE band gaps. The normal PBE method usually underestimates the band gap, but the mBJ method, comparable with GW, usually gets band gap in good agreement with the experimental data.

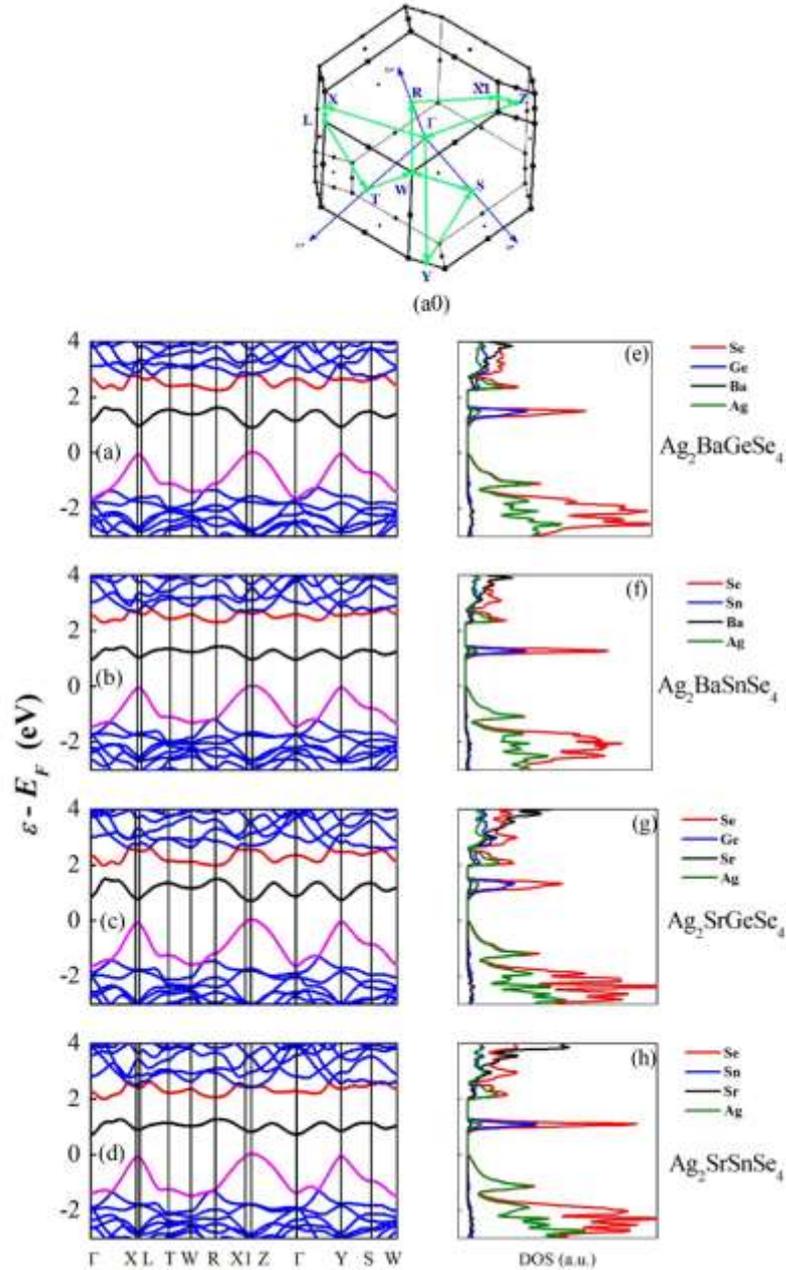

Fig. 2. (Color online) Calculated band structures along high symmetry points showed in (a0) and DOS for $Ag_2XYSe_4$ (X=Ba, Sr; Y=Sn, Ge).

The DOS diagrams show Se-Ag hybridization is the major ingredient of valence band (VB) in all the compounds, which implies Se-Ag covalent bonds are in the four compounds. Conduction band (CB) displays significant density of Se and Ge/Sn atoms. The Ba/Sr atoms have negligible contributions to both VB and CB in each compound. It is implied that doping vacancy or heterogeneous atom at the site of Ba/Sr atoms can result in little effect on electronic transport properties. The valence bandwidth (VBW) is ~2 eV greater than the conduction bandwidth (CBW) of less than 1 eV in each compound. This possibly implies that n-type carrier has larger effective mass than p-type carrier does.

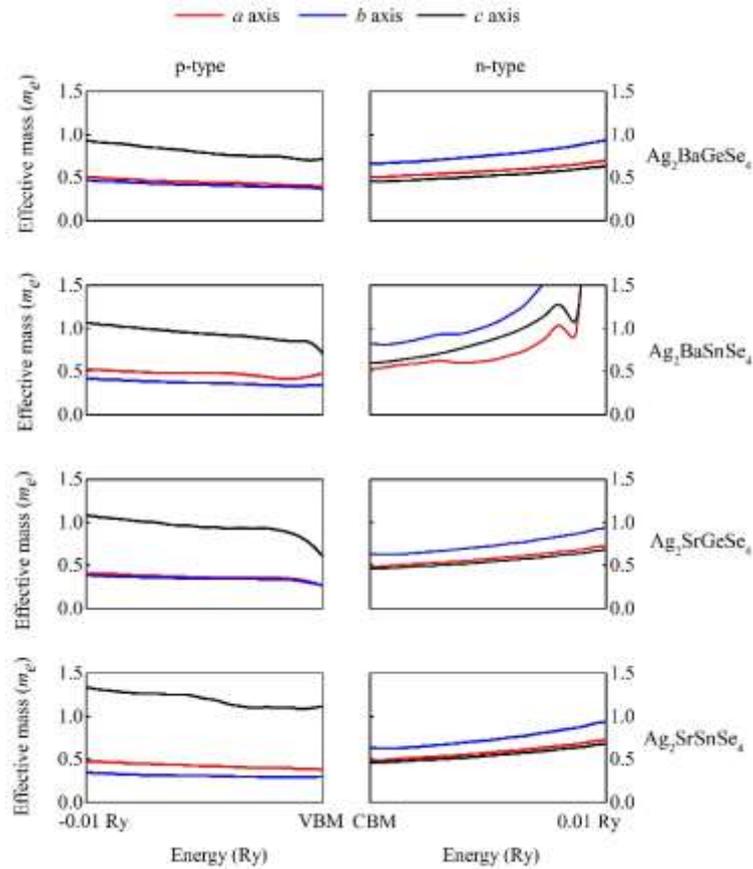

Fig. 3. (Color online) The effective mass of p- and n-type carriers for $Ag_2XYSe_4$ (X=Ba, Sr; Y=Sn, Ge) as a function of energy. Both VBM and CBM are set to 0 eV.

Calculated effective mass along the *a*, *b* and *c* axes as a function of energy is plotted in Fig. 3. It is expect that the electron effective mass is larger than the hole effective mass along the *a* and *b* axes in each compound, although it is in the opposite case along the *c* axis. In addition, the hole effective mass decrease with the increasing energy, which is the opposite of the electron effective mass. For each compound, the hole effective mass along the *c* axis is larger than along the *a* and *b* axes in each compound, and the electron effective mass along the *b* axis is larger than along the other axes. For instance, the effective mass for p-type $Ag_2SrSnSe_4$ along the *c* axis can reach up to 1.33 $m_e$, and the effective mass for n-type $Ag_2BaSnSe_4$ is as large as 9.82 $m_e$ (not shown in Fig. 3) at ~0.01 Ry above the CBM.

TABLE I. Experimental and calculated lattice constants (in Å) and calculated band gaps (in eV) of $Ag_2XYSe_4$ (X=Ba, Sr; Y=Sn, Ge) using different exchange-correlation functionals (PBE, mBJ and HSE06).

|  | $Ag_2BaGeSe_4$ | | $Ag_2BaSnSe_4$ | | $Ag_2SrGeSe_4$ | | $Ag_2SrSnSe_4$ | |
|---|---|---|---|---|---|---|---|---|
|  | This work | Expt.[a] | This work | Expt.[b] | This work | Theo.[c] | This work | Theo.[c] |
| *a* (Å) | 7.053 | 7.058 | 7.141 | 7.116 | 7.110 | 7.115 | 7.183 | 7.193 |
| *b* (Å) | 7.501 | 7.263 | 7.713 | 7.499 | 7.413 | 7.389 | 7.673 | 7.657 |
| *c* (Å) | 8.481 | 8.263 | 8.582 | 8.337 | 8.061 | 7.951 | 8.135 | 8.034 |
| PBE Bandgap (eV) | 0.284 | | 0.195 | | 0.000 | | 0.041 | |
| mBJ Bandgap (eV) | 0.909 | | 0.832 | | 0.708 | | 0.729 | |
| HSE06 Bnadgap (eV)[c] | 0.85 | | 0.77 | | 0.68 | | 0.66 | |

[a] From Ref. [30]
[b] From Ref. [45]
[c] From Ref. [31]

### B Structure stabilities

In order to verify structural stability, phonon spectrum with corresponding phonon DOS for the four compounds was calculated and plotted in Fig. 4. There are eight atoms in one primitive cell and one atom can yield three phonon branches. Thus, there are twenty four phonon branches composed of sixteen longitudinal branches and eight transverse branches, including one longitudinal acoustic (LA), two transverse acoustic (TA) modes.

Phonon DOS in the range of low frequency mainly comes from the contribution of Ag atom. The phonon thermal transport property is primarily decided by phonon modes of low frequency. This implies Ag atom play important role and then Sn atom has negligible effect in lattice thermal transport. As is known, for a system with dynamical stability, the frequency of each phonon mode should be not imaginary. Our result shows no obvious imaginary frequencies in the full Brillouin zone, thus ensuring structure stability.

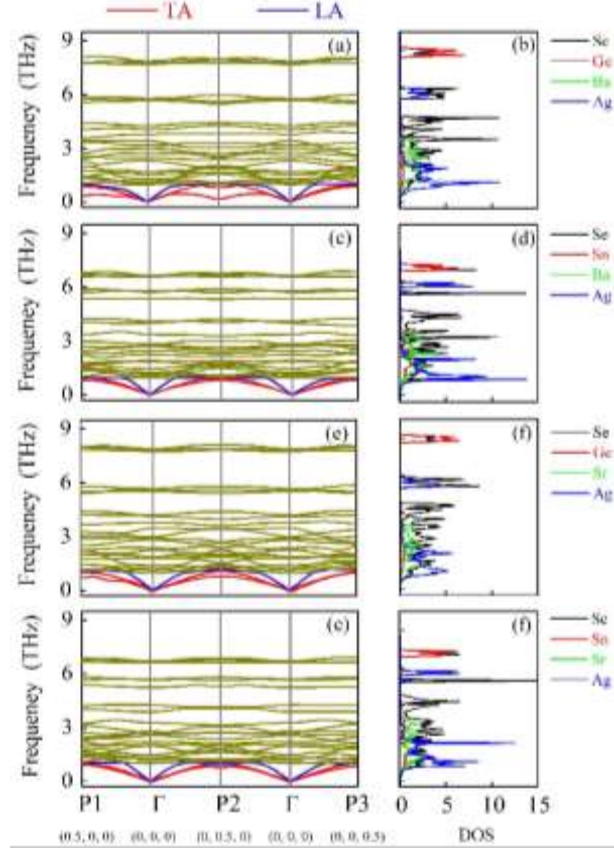

Fig. 4. (Color online) The phonon spectrums along high symmetry points for Ag$_2$XYSe$_4$ (X=Ba, Sr; Y=Sn, Ge) and phonon DOS.

Elastic constants were obtained by second-order-derivative of polynomial fit of energy as a function of strain at zero strain, which called energy approach. For the orthorhombic Ag$_2$XYSe$_4$, there are nine independent elastic constants ($c_{11}$, $c_{22}$, $c_{33}$, $c_{44}$, $c_{55}$, $c_{66}$, $c_{12}$, $c_{13}$ and $c_{23}$). The elastic constants $c_{11}$, $c_{22}$ and $c_{33}$ are indicative of the stiffness against principal strains that are along the *a*, *b* and *c* axes, respectively. The other elastic constants represent resistance against shear deformations, which are obviously smaller than the $c_{ii}$ ($i$=1, 2, and 3). Table II shows the $c_{11}$, $c_{22}$ and $c_{33}$ follow the order of $c_{11} > c_{22} > c_{33}$. Interestingly, the lattice constants follow the opposite order of *a* < *b* < *c*. The elastic constants can be used to assess structural mechanical stability. The criteria of mechanical stability for orthogonal crystals are expressed by [46]

$$c_{11} > 0, c_{22} > 0, c_{33} > 0, c_{44} > 0, c_{55} > 0, c_{66} > 0, \tag{8}$$
$$c_{11} + c_{22} + c_{33} + 2c_{12} + 2c_{13} + 2c_{23} > 0, \tag{9}$$
$$c_{11} + c_{22} - 2c_{12} > 0, \tag{10}$$
$$c_{11} + c_{13} - 2c_{13} > 0, \tag{11}$$
$$c_{22} + c_{33} - 2c_{23} > 0. \tag{12}$$

It is clear that the four compounds are of mechanical stability. Some mechanical and thermal parameters can be calculated by the elastic constants [47]. For examples, using the Voigt-Reuss-Hill approximations bulk and shear moduli B and G can be calculated (summarized in Table II). B for each compound is larger than G. The G/B ratio is usually used to evaluate the brittleness of materials. High G/B ratio is indicative of fragile material. For example, the fragile material α-$SiO_2$ has a high G/B ratio of 1.06 [48]. The G/B ratios for the four compound are all less than 0.5, showing they are unbreakable.

Using B and G Poisson's ratio can be computed:

$$\nu = \frac{3B - 2G}{6B + 2G}. \tag{13}$$

In macroscopic view, Poisson's ratio can estimate the corresponding lateral strain to applied axial strain. Microcosmically, it represents the degree of covalent bond. Low value of Poisson's ratio is indicative of covalent bond; high value represents strong ionic or metallic bonds [43]. Poisson's ratios of $Ag_2XYSe_4$ are lower than 0.367 of silver bulk. This indicates Ag atom in $Ag_2XYSe_4$ participates in covalent bonds, which agrees with results from electronic band structures.

Debye temperature $\Theta$ and acoustic Grüneisen parameter $\gamma_a$ are attained using the following formulas:

$$\gamma_a = \frac{9v_l^2 - 12v_t^2}{2v_l^2 + 4v_t^2}, \tag{14}$$

$$\Theta_D = \frac{h}{k_B}\left(\frac{3n_{tot}N_A\rho}{4\pi M}\right)^{1/3} v_m, \tag{15}$$

where physical constants $h$, $N_A$ are the Plank's constant and Avogadro's number, $M$ and $\rho$ are the total atomic mass in primitive unit cell and the volume density of materials, $v_m$ is the average sound velocity given by the velocities for transverse and longitude waves velocities $v_l$ and $v_s$. The three velocities can be given by the following formulas:

$$v_s = \left(\frac{3B + 4G}{3\rho}\right)^{1/2}, \quad v_p = \left(\frac{G}{\rho}\right)^{1/2}, \tag{16}$$

$$v_m = \left[\frac{1}{3}\left(\frac{2}{v_s^3} + \frac{1}{v_p^3}\right)\right]^{-1/3}, \tag{17}$$

Due to similarities of structure and composition, the four compounds have close values of both $\Theta$ and $\gamma_a$. The largest difference of $\Theta$ is 12.36 K between Ag$_2$BaGeSe$_4$ and Ag$_2$SrSnSe$_4$.

Grüneisen parameter scales the degree of anharmonic vibrations. The Grüneisen parameters $\gamma_a$ of the four compounds are about 1.8, the Ag$_2$SrSnSe$_4$ has the largest $\gamma_a$ value of 1.85, indicating strong anharmonic vibrations of acoustic wave. As is known the thermal resistivity (reciprocal thermal conductivity) origins from anharmonic vibrations of phonon and especially acoustic phonon. It is noted that recent experiential work shows the average Grüneisen parameter of Ag$_2$BaSnSe$_4$ is very small value of 0.65 [49] that is less than the acoustic Grüneisen parameter ~1.8 in this work. This implies that anharmonic vibrations of optical phonon are very weak.

TABLE II. Calculated elastic constants $c_{ij}$ in GPa, bulk and shear moduli $B$ and $G$ in GPa, velocities of transverse and longitudinal waves $v_t$ and $v_l$ in $m/s$, Poisson's ratios $v$, Debye temperatures $\Theta$ in K and acoustic Grüneisen parameters $\gamma_a$ for Ag$_2$XYSe$_4$ (X=Ba, Sr; Y=Sn, Ge).

|  | Ag$_2$BaGeSe$_4$ | Ag$_2$BaSnSe$_4$ | Ag$_2$SrGeSe$_4$ | Ag$_2$SrSnSe$_4$ |
|---|---|---|---|---|
| $c_{11}$ | 90.1953 | 90.8454 | 98.4429 | 99.8721 |
| $c_{22}$ | 56.0873 | 64.6186 | 66.9605 | 63.6739 |
| $c_{33}$ | 52.2170 | 52.8386 | 62.3014 | 56.6002 |
| $c_{44}$ | 24.5931 | 22.7369 | 24.7166 | 23.2939 |
| $c_{55}$ | 26.0964 | 21.4068 | 25.2265 | 23.4543 |
| $c_{66}$ | 33.7824 | 31.6366 | 29.2304 | 30.8721 |
| $c_{12}$ | 48.8688 | 53.5752 | 54.4242 | 56.4909 |
| $c_{13}$ | 33.0205 | 29.9881 | 35.1455 | 32.7599 |
| $c_{23}$ | 9.3571 | 30.3142 | 29.9960 | 30.7913 |
| $B$ | 44.207 | 45.954 | 49.460 | 48.150 |
| $G$ | 20.685 | 19.939 | 21.969 | 20.606 |
| $v_t$ | 1944.30 | 1900.12 | 2016.75 | 1943.25 |
| $v_l$ | 3622.08 | 3624.24 | 3818.37 | 3722.74 |
| $v_a$ | 2171.08 | 2125.24 | 2254.55 | 2174.10 |
| $v$ | 0.297 | 0.310 | 0.306 | 0.312 |
| $\Theta$ | 212.535 | 204.537 | 224.897 | 212.985 |
| $\gamma_a$ | 1.75 | 1.83 | 1.81 | 1.85 |

**C Thermoelectric properties**

1. Carrier relaxation time

It is well known that the precise calculation for the carrier relaxation time is difficult to be implemented due to complex carrier-scattering mechanism. In this work, we calculated the relaxation time based on considering the carrier-phonon scattering using the Fermi-golden rule and the deformation potential (DP) theory. According to the Fermi-golden rule, the scattering rate $P_k$ at the electronic state $k$ can be described by

$$P_{kk'} = \frac{2\pi}{\hbar} \sum_{k'} |M(k,k')|^2 \delta[\varepsilon(k) - \varepsilon(k') \pm \Delta E], \tag{18}$$

where the matrix element M($k$, $k'$) represents the scattering from $k$ state to $k'$ state, which is known as the transition matrix element, $\Delta E$ is the phonon energy. In this work, we taken the phonon energy as $3k_BT$. Based on solid state physics theory, the reciprocal of the carrier relaxation time can expressed as:

$$\frac{1}{\tau(k)} = \sum_{k'} P_{kk'}(1-\cos\theta) = \frac{2\pi}{\hbar} \sum_{k'} |M(k,k')|^2 \delta[\varepsilon(k) - \varepsilon(k') \pm \Delta E](1-\cos\theta). \tag{19}$$

Herein, the angle $\theta$ is between the $k$ state to the $k'$ state. In the scattering process, the carrier absorbs or emits a phonon. Thus, by the DP theory, the effective transition matrix element around the valence band maximum (VBM) or conduction band maximum (CBM) can be given by [50]

$$|M|^2 = \frac{k_B T E_\beta^2}{\Omega c_\beta}, \tag{20}$$

where $E_\beta$ and $c_\beta$ are the deformation potential constant and elastic constant along the $\beta$ direction. It is noted that the anisotropy is neglected and the matrix element $M$ is independent of the $\theta$ in the DP theory, and thus the above equation can be described as

$$\frac{1}{\tau(k)} = \frac{2\pi}{\hbar} \sum_{k'} |M(k,k')|^2 \delta[\varepsilon(k) - \varepsilon(k') \pm \Delta E], \tag{21}$$

it is worth mentioning that the anisotropy is not completely ignored in the DP theory, and it arises from the anisotropic deformation potential constants and anisotropic elastic constants, which has small difference in the different directions.

The carrier relaxation time was showed in Fig. 5. It is found that the $\tau$ for each compound reduces with increasing $T$ and is of the order of magnitude of $10^{-14} \sim 10^{-15}$. The $\tau$ for p-type $Ag_2BaSnSe_4$ and $Ag_2SrSnSe_4$ is obviously higher than their n-type. However, for $Ag_2BaGeSe_4$ and $Ag_2SrGeSe_4$, p- and n-type carriers have close values of relaxation time. It is worth mentioning that the scatters from kinds of defects were not considered. Thus, the actual value should be lower that the calculated.

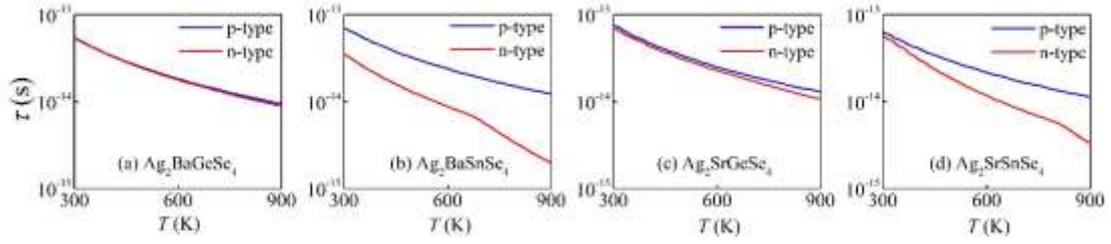

Fig. 5. (Color online) The relaxation time of p- and n-type carrier as a function of temperature for $Ag_2XYSe_4$ (X=Ba, Sr; Y=Sn, Ge).

2. Electronic transport properties

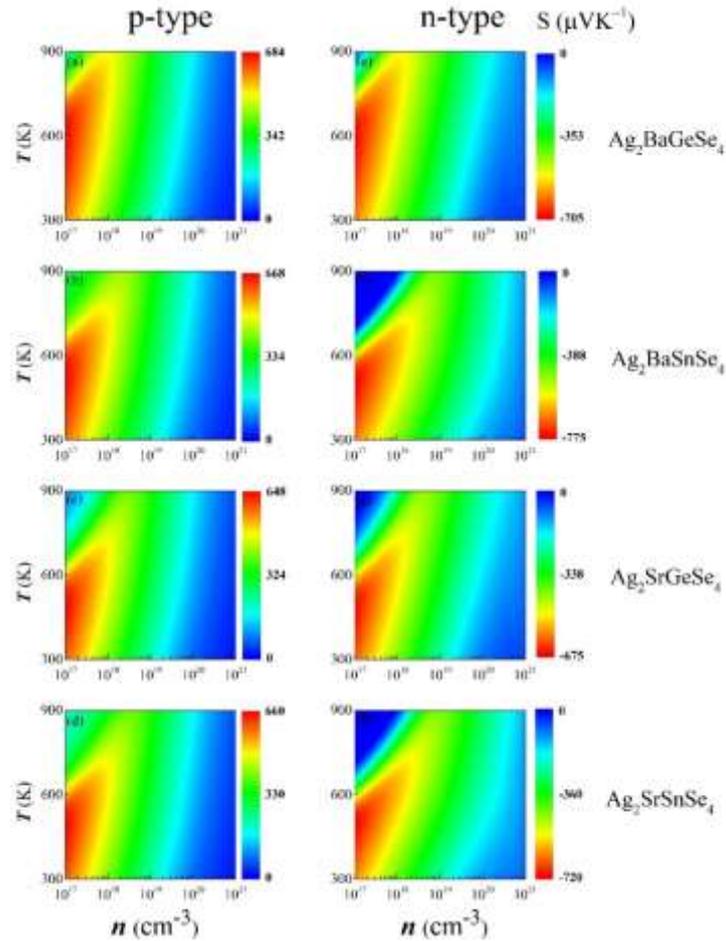

Fig. 6. (Color online) Calculated Seebeck coefficient with respect to temperature $T$ and carrier density $n$ for p- and n-type $Ag_2XYSe_4$ (X=Ba, Sr; Y=Sn, Ge).

According to Boltzman theory, Seebeck coefficient $S$, electrical conductivity $\sigma$ and electronic thermal conductivity $\kappa_e$ are all related to temperature and chemical potential. The chemical potential corresponds to carrier density. Thus, calculated $S$ with respect to temperature $T$ and carrier density $n$ for p-and n-type Ag$_2$XYSe$_4$ (X=Ba, Sr; Y=Sn, Ge) were calculated and then plotted in Figs. 6. The largest |S| of n-type system is higher than that of the p-type system. For example, the largest |S| of n- and p-type Ag$_2$BaSnSe$_4$ are 775 and 668 μeV respectively. The largest |S| for each n-type systems is above 700 μeV. |S| follows the following rules: 1. The |S| reduces with increasing carrier density at low temperatures, and reduces and then rises at high temperature. 2. The |S| rises and then reduces with increasing temperature at low carrier density. The |S| rises with increasing temperature at high carrier density.

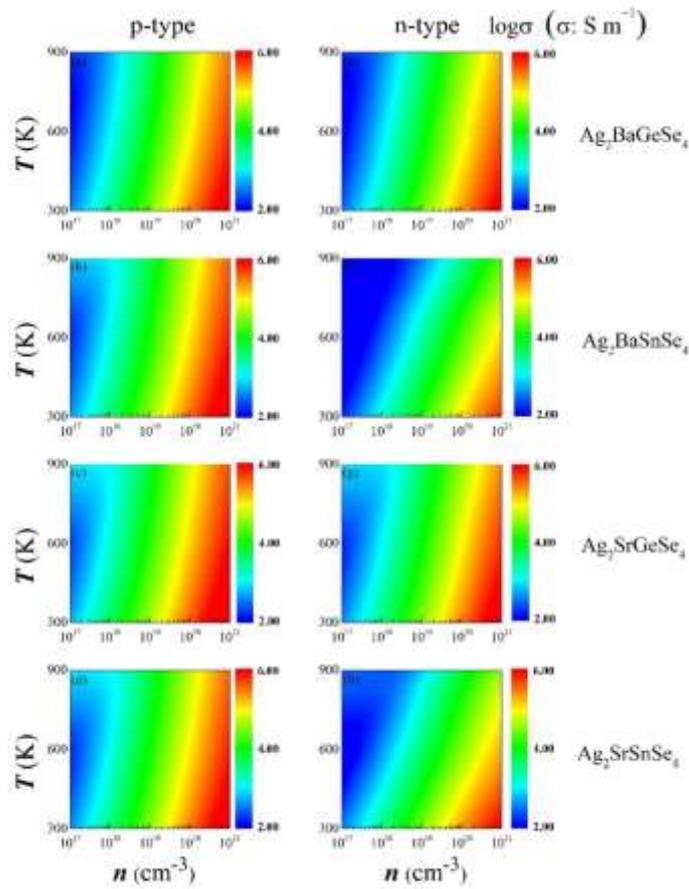

Fig. 7. (Color online) Calculated electrical conductivity with respect to temperature $T$ and carrier density $n$ for p- and n-type Ag$_2$XYSe$_4$ (X=Ba, Sr; Y=Sn, Ge).

Fig. 7 shows that the electric conductivity is more sensitive to carrier density than to temperature. In the case of p-type Ag$_2$BaSnSe$_4$, $\sigma$ at 300 K changes from 4.12 Scm$^{-1}$ to 2.71×10$^4$ Scm$^{-1}$ with the carrier density increasing from 10$^{17}$ to 10$^{21}$ cm$^{-3}$. $\sigma$ at 300 K reduces to 1.46 Scm$^{-1}$ at 620 K and then rises to 7.33 Scm$^{-1}$ at 900 K. Fig. 8 shows that electronic thermal conductivity at low carrier densities < 10$^{19}$ cm$^{-3}$ has negligible contribution to the total thermal conductivity. At 300 K, $\kappa_e$ of p-type Ag$_2$BaGeSe4 is 0.004 Wm$^{-1}$K$^{-1}$ at $10^{17}$ cm$^{-3}$, which can reach up to 0.37 Wm$^{-1}$K$^{-1}$ at $10^{19}$ cm$^{-3}$. $\kappa_e$ is not sensitive to the change of temperature like $\sigma$. At $10^{19}$ cm$^{-3}$, when temperature increases 300 K to 900 K, $\kappa_e$ reduces to 0.06 Wm$^{-1}$K$^{-1}$.

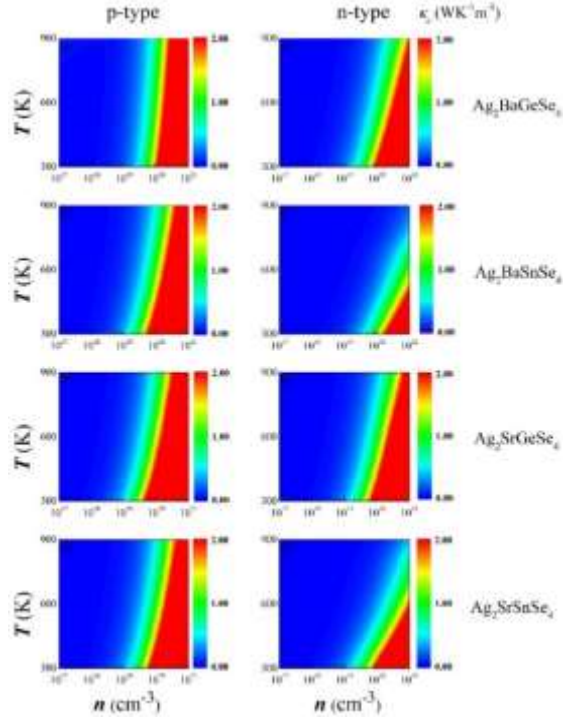

Fig. 8. (Color online) Calculated electronic thermal conductivity with respect to temperature $T$ and carrier density $n$ for p- and n-type Ag$_2$XYSe$_4$ (X=Ba, Sr; Y=Sn, Ge).

3. Phonon transport properties

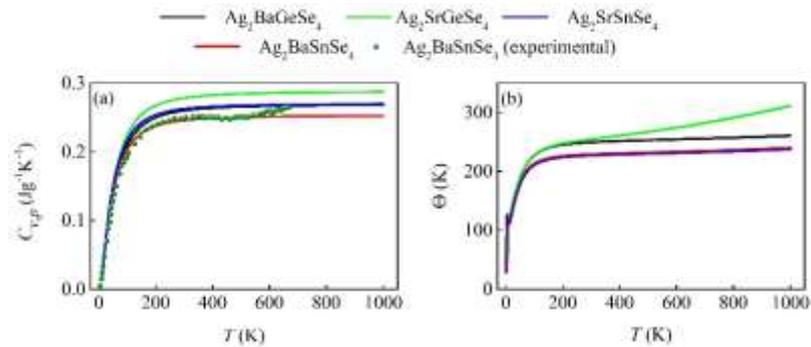

Fig. 9. (Color online) (a) Calculated heat capacity at constant volume for Ag$_2$XYSe$_4$ (X=Ba, Sr; Y=Sn, Ge) and experimental heat capacity for Ag$_2$BaSnSe$_4$, (b) Calculate Debye temperature for Ag$_2$XYSe$_4$ (X=Ba, Sr; Y=Sn, Ge)

Fig. 9 shows the calculated specific heat at constant volume $C_v$ of the four compounds and experimental specific heat at constant press $C_p$ of $Ag_2BaSnSe_4$. Below 550 K, $C_v$ and $C_p$ of $Ag_2BaSnSe_4$ agree with each other. Above 550 K, $C_p$ is higher than $C_v$. The difference between $C_v$ and $C_p$ above 550 K origins from the thermal expansion coefficient and isothermal compressibility, which is reasonable and acceptable. $\Theta$ was also obtained by fitting the calculated specific heat $C_v$ according to the following equation:

$$C_V = 9Nk_B \left(\frac{T}{\Theta_D}\right)^3 \int_0^{\Theta_D/T} \frac{x^4 e^x}{\left(e^x-1\right)^2} dx, \quad (22)$$

herein, N represents the total number of atoms in the primitive cell. In Fig. 9(b), $\Theta$ at 300 follows the order of $Ag_2SrGeSe_4$ (255.0K) > $Ag_2BaGeSe_4$ (249.9) > $Ag_2BaSnSe_4$ (228.3) > $Ag_2SrSnSe_4$ (228.0), which indicates the values are slightly higher than these by Eq. (15). Experimental $\Theta$ of $Ag_2BaSnSe_4$ is about 160 K [49] that is obviously lower than the calculated value of 244.9 K. We consider the different fitting equations lead to the difference of Debye temperatures between in this work and Ref. [49].

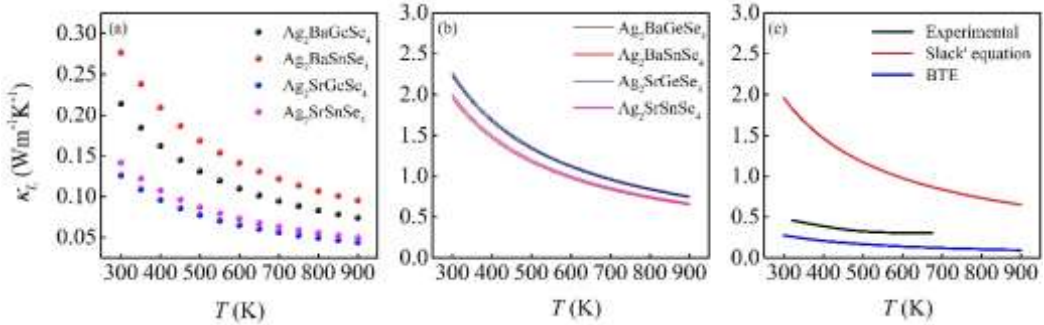

Fig. 10. (Color online) Calculated lattice thermal conductivity obtained by (a) BTE and (b) Slack's equation for for $Ag_2XYSe_4$ (X=Ba, Sr; Y=Sn, Ge), (c) comparisons of the calculated and experimental lattice thermal conductivities for $Ag_2BaSnSe_4$.

We used the slack's equation and BTE to calculate the lattice thermal conductivity of $Ag_2XYSe_4$, respectively. Lattice thermal conductivities calculated by two equations both reduce with rising temperatures (see Fig. 10 (a) and (b)). $\kappa_L$ obtained by slack's equation is higher than that by BTE. $\kappa_L$ of $Ag_2BaSnSe_4$ by BTE is more close to experimental data (see Fig. 10 (c)) from Ref. [49] but that by the Slack's equation seems more reasonable. The calculated $\kappa_L$ is without the effects of grain boundary, vacancy and doping, and only is average of anisotropic property, which should be higher than experimental data of materials containing all kinds of defects. Thus, we used $\kappa_L$ by the Slack's equation to calculated ZT.

4. Figure of merit

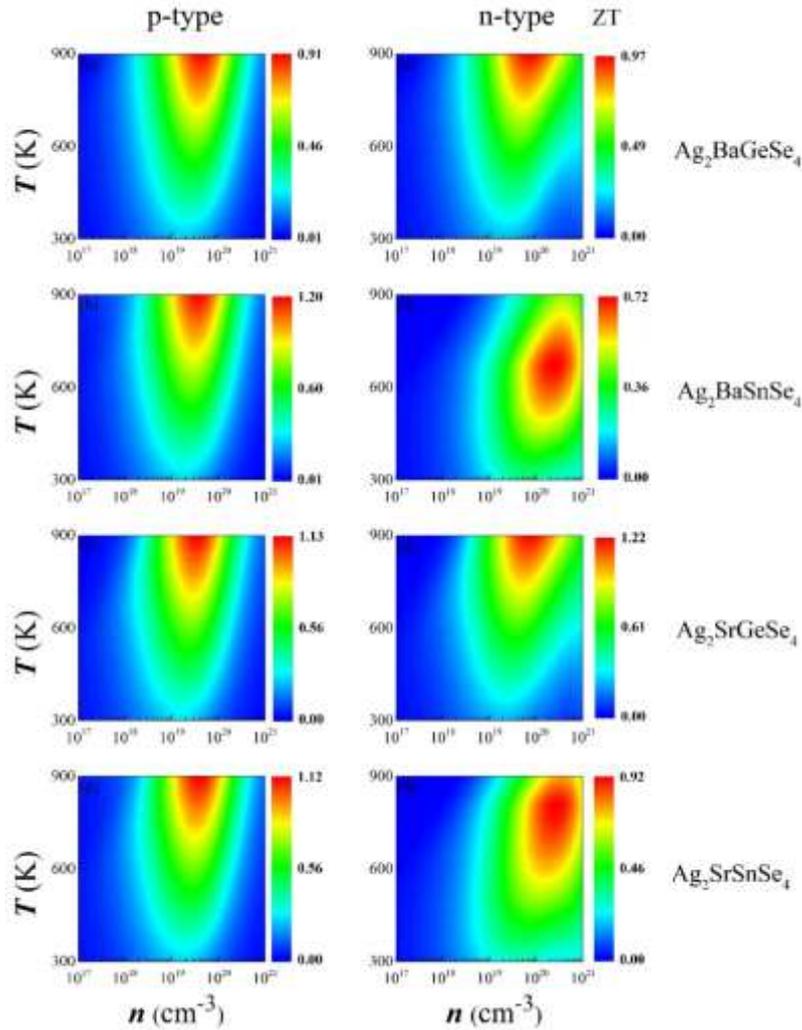

Fig. 11. (Color online) Calculated ZT with respect to temperature $T$ and carrier density $n$ for p- and n-type $Ag_2XYSe_4$ (X=Ba, Sr; Y=Sn, Ge).

ZT with respect to $T$ and $n$ in a sufficiently broad region ($n$, $T$) were plotted in Fig. 11, where the red, green and blue areas denote the high, moderate, and low ZT. N-type $Ag_2BaSnSe_4$ and Ag2SrSnSe4 possess respectively low ZT maximums 0.72 and 0.92, although they have larger red closed region than the other systems have the red open region. ZT maximums for p- and n-type $Ag_2SrGeSe_4$ reach up to 1.22 and 1.13, respectively. However, these for p- and n-type $Ag_2BaGeSe_4$ are below unit (0.91 and 0.97). ZT maximums for P-type $Ag_2BaSnSe_4$ and $Ag_2SrSnSe_4$ are 1.20 and 1.12, respectively. For all the p-type systems and n-type $Ag_2BaGeSe_4$ and $Ag_2SrGeSe_4$, ZT maximums appear at 900 K and in the range of $10^{19} \sim 10^{20}$ cm$^{-3}$. However, these for n-type $Ag_2BaSnSe_4$ and $Ag_2SrSnSe_4$ appear in the range of $10^{20} \sim 10^{21}$ cm$^{-3}$ and respectively at 670 K and 810 K.

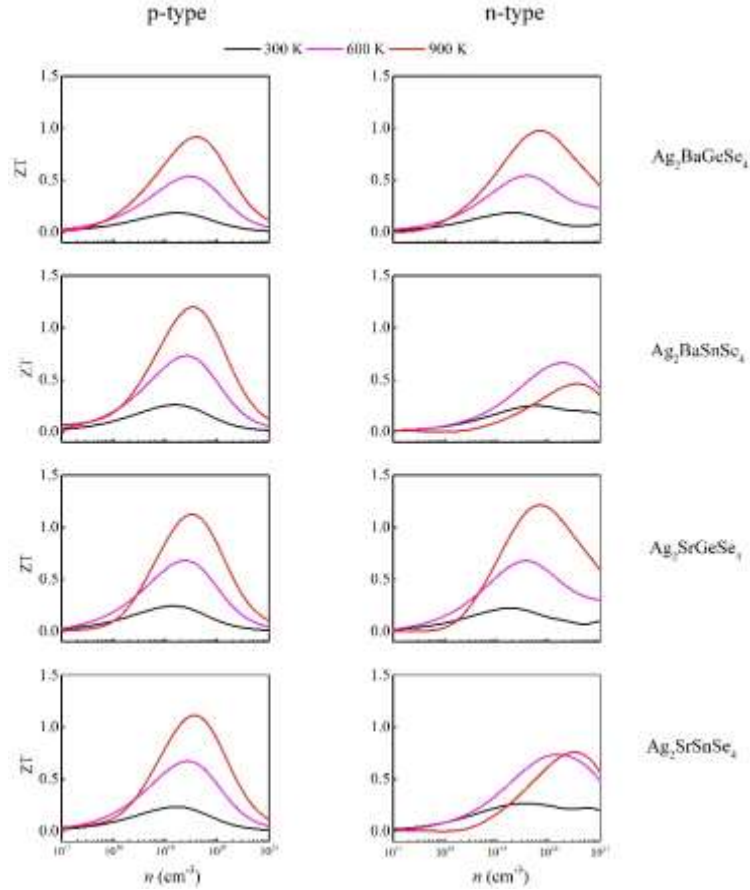

Fig. 12. (Color online) Calculated ZT as a function of carrier density at 300 K, 600 K and 900 K for p- and n-type $Ag_2XYSe_4$ (X=Ba, Sr; Y=Sn, Ge).

In order to study TE properties in detail, we extract ZT as a function of carrier density $n$ at 300 K, 600 K and 900 K from Fig. 11 and they were showed in Fig. 12. For all the systems, ZT rises to the peak value and then reduces with increasing carrier density. At most of carrier densities, ZT rises with increasing temperature. In the range of $10^{18} \sim 10^{19}$ cm$^{-3}$, ZT is sensitive to both carrier density and temperature. At high temperature, ZT is more sensitive to carrier density. In the case of p-type $Ag_2BaSnSe_4$, The ZT at 900 K changes from 0.06 to 1.20 with the carrier density increasing from $10^{17}$ cm$^{-3}$ to $3.3 \times 10^{19}$ cm$^{-3}$.

In this work, we defined the optimal carrier density $n_{opt}$ and the optimal temperature $T_{opt}$ at which the system has the ZT maximum. The $n_{opt}$ and $T_{opt}$ of the four compounds with p- and n-type carrier were summaried in Table III. The $n_{opt}$ for most systems are ~$10^{19}$ cm$^{-3}$, but $n_{opt}$ for n-type Ag$_2$BaSnSe$_4$ and Ag$_2$SrSnSe$_4$ reach up to $2.5 \times 10^{20}$ cm$^{-3}$ and $2.7 \times 10^{20}$ cm$^{-3}$, respectively. The optimal carrier density is very useful for promoting ZT by experimental doping. Theoretically, one can easily calculate the optimal carrier density, but the optimal carrier density and the optimal temperature are difficult to achieve experimentally. For example, when temperature rises to 900 K, material may undergo phase transition or melt. These are the issues to be considered and solved in the experiment, but we still consider Ag$_2$XYSe$_4$ (X=Ba, Sr; Y=Sn, Ge) as a class of potential thermoelectric materials.

TABLE III Optimal carrier density and optimal temperature for p- and n-type Ag$_2$XYSe$_4$ (X=Ba, Sr; Y=Sn, Ge).

| Compounds | p-type | | | n-type | | |
|---|---|---|---|---|---|---|
| | $n_{opt}$ (cm$^{-3}$) | $T_{opt}$ (K) | ZT$_{max}$ | $n_{opt}$ (cm$^{-3}$) | $T_{opt}$ (K) | ZT$_{max}$ |
| Ag$_2$BaGeS$_4$ | $4.3 \times 10^{19}$ | 900 | 0.91 | $6.9 \times 10^{19}$ | 900 | 0.97 |
| Ag$_2$BaSnSe$_4$ | $4.0 \times 10^{19}$ | 900 | 1.20 | $2.5 \times 10^{20}$ | 670 | 0.72 |
| Ag$_2$SrGeSe$_4$ | $3.3 \times 10^{19}$ | 900 | 1.13 | $6.9 \times 10^{19}$ | 900 | 1.22 |
| Ag$_2$SrSnSe$_4$ | $3.7 \times 10^{19}$ | 900 | 1.12 | $2.7 \times 10^{20}$ | 810 | 0.92 |

## IV. CONCLUSIONS

In summary, we use the first-principles method to systematically explore electronic structures, mechanical and thermal and TE properties of p- and n-type Ag$_2$XYSe$_4$ (X=Ba, Sr; Y=Sn, Ge). The results show that low crystal symmetry and suitable band gap mainly contribute to weak thermal transport and medium electrical transport performances. The largest ZT in the p-type Ag$_2$XYSe$_4$ comes from Ag$_2$BaSnSe$_4$ compound reaching up to 1.20 at 900 K, and the largest ZT in the n-type is 1.22 of Ag$_2$SrGeSe$_4$ at 900 K. Although the values for ZT is not very excellent, we believe the ZT for those compounds can be prompted by such as doping, low-dimensional strategy.

**ACKNOWLEDGMENTS**

Authors gratefully acknowledge great support from the National Natural Science Foundation of China (Grant No. 11804132).